\documentclass[11pt]{article}
\usepackage{hyperref}
\usepackage{graphicx}
\pdfoutput=1
\begin{document}

\title{Real Time Electron Microscope Imaging of \\ Nanoparticle Motion Induced by a \\ Moving Contact Line}

\author{Joseph M. Grogan and Haim H. Bau \\
\\\vspace{6pt} Department of Mechanical Engineering and Applied Mechanics, \\ University of Pennsylvania, Philadelphia, PA 19104, USA}

\maketitle

\begin{abstract}
Real time fluid dynamics videos showing the motion and aggregation of nanorods induced by moving contact lines were obtained by liquid cell {\it in situ} scanning transmission electron microscopy.
\end{abstract}

\section{Main Text}

\begin{figure}[h!]
  \centering
    \includegraphics[width=0.9\textwidth]{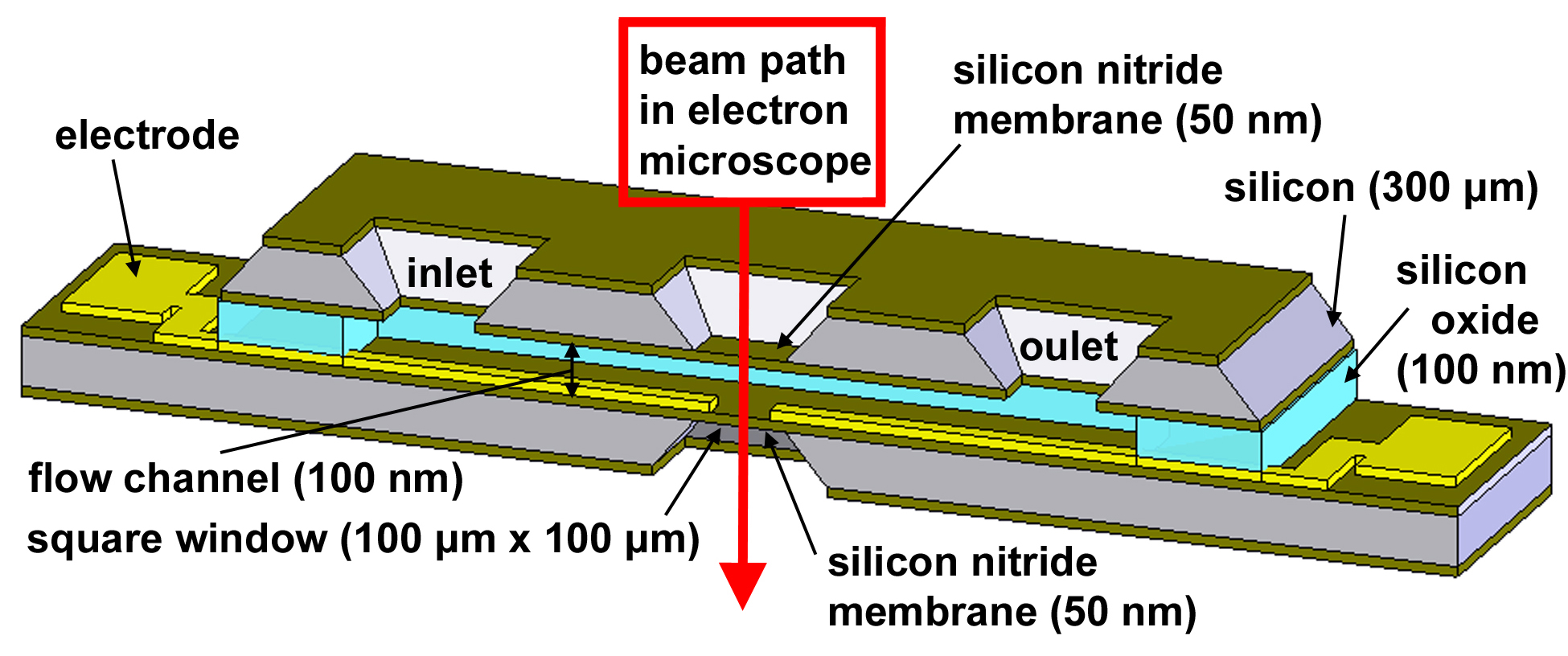}
  \caption{\small A schematic depiction of the nanoaquarium's cross section. Not drawn to scale.\label{fig:nanoaquarium}}
\end{figure}

The nanoaquarium is a custom-made microfabricated nanofluidic device that enables transmission electron microscope (TEM) imaging of dynamical processes taking place in liquid media \cite{GroganBauJMEMS,GroganBauPRE}. The nanoaquarium (Figure~\ref{fig:nanoaquarium}) consists of a shallow conduit ($\sim$ 100 nm) sandwiched between two thin silicon nitride membranes (each 50 nm thick). The liquid in the device is hermetically sealed and isolated from the high vacuum environment of the microscope chamber to prevent evaporation. The conduit height and membrane thickness are among the thinnest of any liquid cell TEM device, enabling the nanoaquarium to yield high resolution, high contrast images with minimal electron scattering by the suspending medium. Electrodes are integrated into the imaging chamber for {\it in situ} actuation and sensing.

\begin{figure}[h!]
  \centering
    \includegraphics[width=0.9\textwidth]{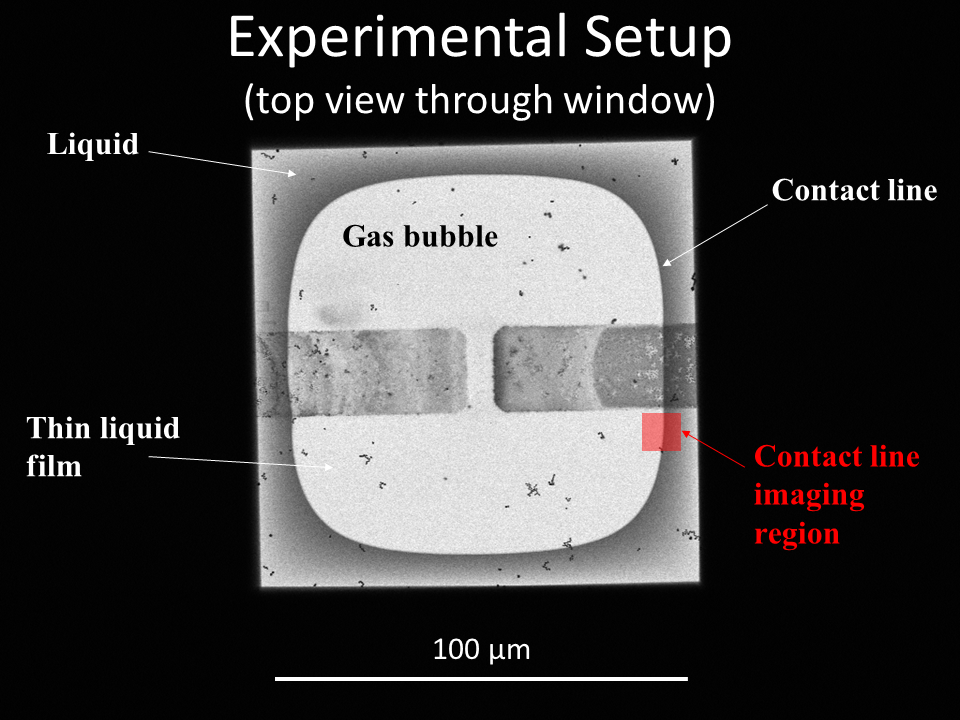}
  \caption{\small Scanning transmission electron microscope image of the nanoaquarium with a bubble (light gray) occupying most of the imaging window and liquid (dark gray) around the perimeter.\label{fig:bubble}}
\end{figure}

The nanoaquarium was filled with an aqueous suspension of gold nanorods (20 nm in diameter,  40 nm in length) stabilized with surfactant cetrimonium bromide (CTAB), a gift from Xingchen Ye and Professor Christopher B. Murray (Chemistry, Penn). Imaging was performed in a FEI Quanta 600 FEG Mark II with a STEM (scanning TEM) detector, operated at 30 kV. Electrical potential of $\sim$ 15 V was applied across the electrodes to generate a bubble that displaced liquid (and nanorods) to the perimeter of the observations chamber. We focused our observations at the interface between the bubble and the `bulk' liquid around the perimeter of the imaging window, which we refer to as the contact line (Figure~\ref{fig:bubble}). The term contact line is used here to denote a narrow region in which the thickness of the liquid film increases rapidly \cite{degennes,Pham}. When the electron beam of the microscope was zoomed in on the interface, movement of the contact line resulted. Sometimes the contact line receded, sometimes it advanced, and sometimes it oscillated back and forth. Fluid dynamics videos were recorded of the interactions between nanorods and the moving contact line.

\begin{figure}[h!]
  \centering
    \includegraphics[width=0.9\textwidth]{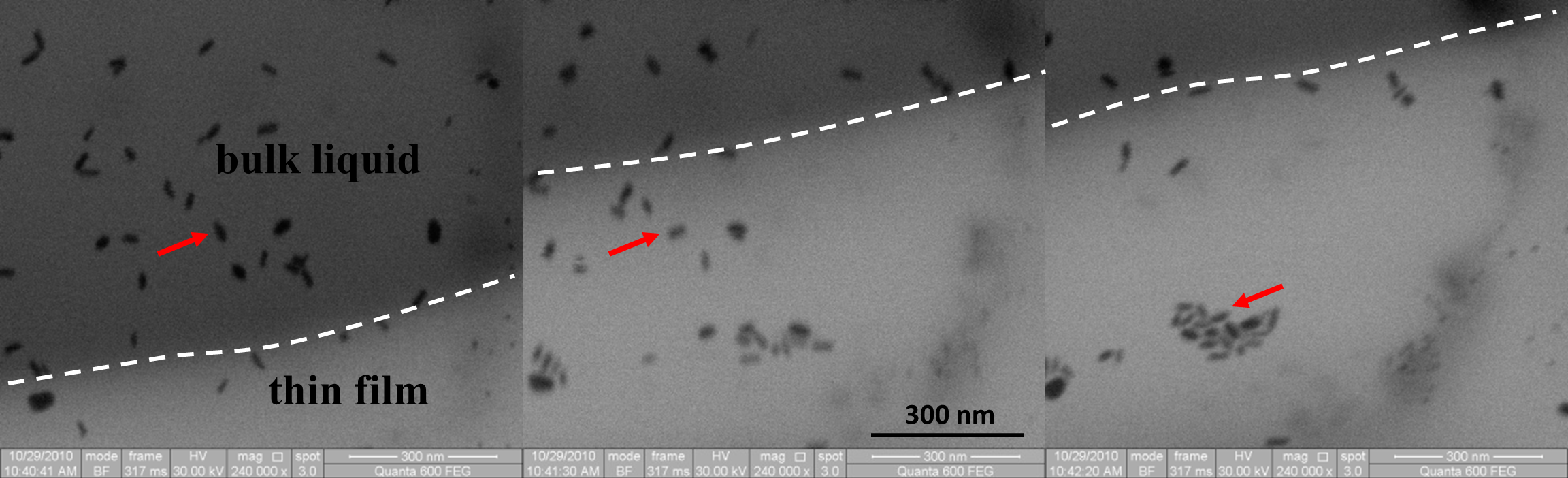}
  \caption{\small Gold nanorods aggregate on a surface as they are ejected from a receding liquid contact line. The dark region moving upward in the three images is the receding liquid and the contact line is indicated by a dashed white line. The same particle is indicated by a red arrow in each frame. 50 seconds elapse between frames. The indicated particle does not move from its initial position on the surface until the contact line has passed over it by some distance.\label{fig:particles}}
\end{figure}

When the contact line receded (see the {\it in situ} STEM video), nanorods were propelled from the `bulk' liquid into the thin liquid film in the gas/vapor region (opposite from the direction of contact line movement). Interestingly, initially stationary particles did not move significantly until the contact line had passed by the particles. In most cases it was not until the contact line receded past the particle by 10s of nm that the particle shot forward (Figure~\ref{fig:particles}). The {\it in situ} STEM video shows the effect of an advancing contact line on nanorod alignment and aggregation. Surprisingly, the initially stationary nanorods are not engulfed by the advancing contact line but are instead pushed ahead. The nanorods align themselves to form a line that is parallel to the moving interface. The {\it in situ} STEM video also shows nanorod motion and aggregation induced by an oscillating contact line that is moving back and forth. The experimental observations are consistent with a mathematical model that determines the fluid velocity in the thin film by considering the surface tension force and disjoining pressure in the thin liquid film \cite{Groganthesis}.

\vspace{10 mm}

{\em Conclusion:} A liquid cell that enables electron microscope imaging of processes in liquid media was designed, constructed, and tested. The liquid cell was used, for the first time, to image the motion and aggregation of nanorods induced by moving contact lines.

\vspace{10 mm}

{\em Acknowledgement:}
This work was supported, in part, by the National Science Foundation grant \# 1066573

\end{document}